# Twitter should now be referred to as X: How academics, journals and publishers need to make the nomenclatural transition

Jaime A. Teixeira da Silva[1*], Serhii Nazarovets[2*]

[1] Independent researcher, Ikenobe 3011-2, Kagawa-ken, 761-0799, Japan; jaimetex@yahoo.com

[2] Borys Grinchenko Kyiv Metropolitan University, 18/2 Bulvarno-Kudriavska Str., 04053 Kyiv, Ukraine; serhii.nazarovets@gmail.com

* Co-corresponding authors

**Abstract**

Here, we note how academics, journals and publishers should no longer refer to the social media platform Twitter as such, rather as X. Relying on Google Scholar, we found 16 examples of papers published in the last months of 2023 – essentially during the transition period between Twitter and X – that used Twitter and X, but in different ways. Unlike that transition period in which the binary Twitter/X could have been used in academic papers, we suggest that papers should no longer refer to Twitter as Twitter, but only as X, except for historical studies about that social media platform, because such use would be factually incorrect.

Twitter became a dominant social media platform since its launch in 2006, and has been widely used in academic papers as an Altmetric tool to assess an endless variety of responses, perceptions and other social responses to an equally diverse range of topics (Said et al., 2019; Yu et al., 2019; Stokel-Walker, 2023). A search in Scopus for peer-reviewed publications from 2006-2024 whose titles, abstracts, or keywords contain mentions of Twitter found 54,107 documents. Moreover, since 2006, the number of scientific publications related to Twitter has been growing rapidly every year (Fig. 1), which indicates a constant interest in this social network by the academic community. Despite this, we were unable to identify any rules related to specifically how Twitter should be used in scientific publications, including, more worrisomely, the lack of data preservation protocols in place in order for readers and researchers to verify the content of tweets that were used, such as in big data social experiments involving hundreds or thousands of tweets (Teixeira da Silva & Nazarovets, 2023). As illustrated in Figure 1, Twitter continues to be utilized in academic articles.

As noted on the Wikipedia pages describing this process, here and here, the name and brand of Twitter has transitioned to "X", including the replacement of the blue bird symbol by a black-and-white X. However, science is neither stationary, nor static, so research related to Twitter and employing this social media Altmetric tool continues to progress, even though this was a multi-month transition period in 2023. An initial search on Google Scholar (5 November 2023) using the search terms "Twitter X" restricted to 2023 revealed a few results of interest that we debate briefly (Table 1). We chose this search engine because, unlike citation databases, such as Scopus, which only allow for a search in the metadata of publications, Google Scholar allows a search of the full text as well, which is important for the purposes of this study.

Although there was – understandably – some variation in the way in which authors referred to Twitter or X during this period, we found the approach by Lee et al. (DOI 10.3389/fpubh.2023.1280658) to be most consistent and prudent, as Twitter/X, thereby making both brand names visible to readers. Now that "X" has become more widely used and dominant brand name, and given that Twitter Inc. has formally become part of X Holdings, we believe that authors should no longer use Twitter moving forward – even if they think that readers would be able to better identify with that established brand name – because doing so would imply the use of an incorrect brand or product name. The greater risk is that a factual error may be subjected to a post-publication correction. Peer reviewers and editors who oversee research involving X also need to pay attention to authors' inadvertent use of Twitter, correcting this to X. Thus, in our opinion, papers published in 2024 should only refer to pre-Twitter as X, unless they were completing a historical analysis of tweets in the pre-X period. In addition to the transition of the use of the term Twitter to X, researchers using this social media tool will also need to transition the use of the term "tweet" to "post", accordingly.

As social media-based research advances, if authors refer to X as Twitter (except for historical uses), should such papers be subjected to a correction?

**Compliance with Ethical Standards**

**Funding**
This research was not funded.

**Conflicts of interest**
The authors declare no relevant conflicts of interest.

**Ethical approval**
Not applicable.

**Informed consent**
Not applicable.

**Author contributions**
Except for the database search in Scopus, which was conducted by the second author, the authors contributed equally to all other aspects of the paper, including, but not exclusively limited to, conceptual design, discussion, methodology, analysis and validation, writing and editing all versions of the manuscript.

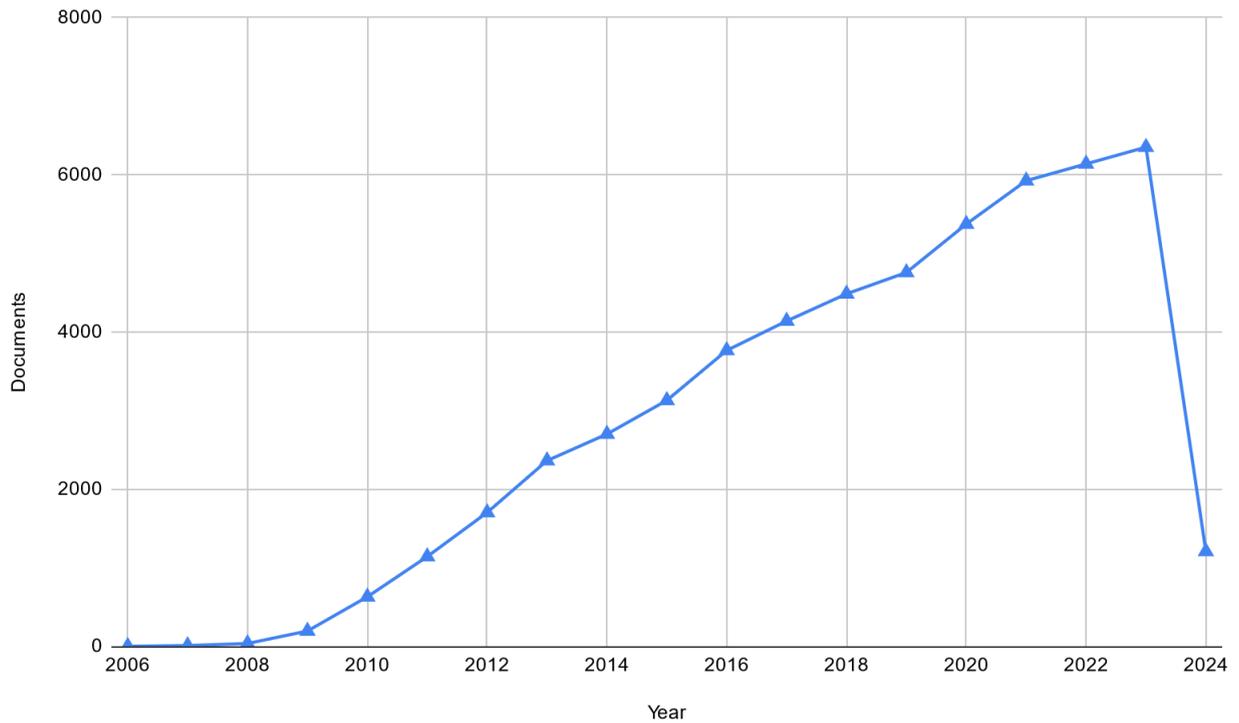

Figure 1 The number of peer-reviewed publications with titles, abstracts, or keywords that mention Twitter in 2006-2024* (Scopus data: retrieved 25 March 2024). Most Twitter-related papers were published in the following areas: Computer Science (32.5%), Social Sciences (17.3%), Engineering (10.5%), Mathematics (7%), Medicine (5.8%), Decision Sciences (5.1%), Arts and Humanities (4.5%), and Business, Management and Accounting (4.3%). * Data for 2024 are incomplete

Table 1 Examples of DOI-indexed studies[1] in which both Twitter (outgoing brand name) and X (incoming brand name) have been used, and the manner in which these have been represented.

| Paper DOI | Manner in which Twitter or X is referenced (verbatim text) |
| --- | --- |
| https://doi.org/10.1080/02640414.2023.2259723 | X (formerly Twitter) (title, abstract) |
| https://doi.org/10.1016/j.cities.2023.104595 | Twitter (title); Twitter (X) (abstract); Twitter (now X) (main text) |
| https://doi.org/10.3389/fpubh.2023.1280658 (OA) | Twitter/X (title, abstract, main text, figures, tables) |
| https://doi.org/10.1108/OXAN-DB281631 | Twitter/X (title; abstract, main text) |
| https://doi.org/10.1016/j.crpvbd.2023.100138 (OA) | Twitter (title, highlights, main text, figures, tables); Twitter/x (once in text); Twitter (X as it is now known) (once in text) |
| https://doi.org/10.3389/fcomm.2023.1264373 (OA) | Twitter (title, abstract, main text); Twitter (X) (once in main text) [2] |
| https://doi.org/10.3390/socsci12110590 (OA) | Twitter (title, abstract, main text, figures); Twitter (now called "X") (once in abstract) [3] |
| https://doi.org/10.48550/arXiv.2309.03064 (OA, preprint) | Twitter (title, main text, tables); Twitter (now X) (abstract, main text) |
| https://doi.org/10.1177/13548565231199981 (OA) | Twitter (title, abstract, main text); Twitter/X (main text); X (main text) |
| https://doi.org/10.1038/s41591-023-02530-1 | twitter [sic] (title); Twitter (main text); Twitter (now known as 'X') (main text) |
| https://doi.org/10.5339/jist.2023.11 (OA) | Twitter (title, abstract, main text); Twitter/X (main text); X (main text) [4] |
| https://doi.org/10.1101/2023.09.17.558161 (OA, preprint) | Twitter (title, abstract, main text); Twitter (now called X) (main text) |
| https://doi.org/10.58729/1941-6679.1561 (OA) | Twitter (title, abstract, main text, tables) [5] |
| https://doi.org/10.1080/15332985.2023.2267712 | Twitter (title, abstract, main text); Twitter (now known as X) (main text) |
| **Non-compliant** | |
| https://doi.org/10.3390/fi15100346 (OA) | Twitter [6] |

Abbreviations: DOI, digital object identifier; OA, open access

Notes:
[1] Non-exhaustive list, discovered using a Google Scholar search (5 November 2023); limited to the first 60 hits
[2] A clear explanation was provided by the authors for their choice of brand name: "Twitter was recently rebranded as "X," however in this article we refer to "Twitter" and "tweets" because data were collected between March and May 2022, prior to the rebranding in July 2023"
[3] Footnote 1 of the article notes: "Towards the end of July 2023, Twitter was rebranded as "X" and a capital letter 'x' substituted the iconic blue bird logo that had represented it since its origins."
[4] The authors note in section 2.4: "X will be referred to Twitter in this section, as the studies were completed before the Elon Musk purchase." It also notes in section 3: "X (formally Twitter) refers to tweets as posts now."
[5] Note on page 66: "Since the name change from Twitter to X, the package and Twitter API continue to be functional and available as of October 2023"
[6] Even though the submission, revision and acceptance dates are indicated as September, October and October 2023, respectively, the authors fail to recognize, or make a note, of the transition in brand name from Twitter to X (a potential case of brand loyalty or resistance to the new brand name).